\documentclass[conference]{IEEEtran}
\IEEEoverridecommandlockouts

\usepackage{cite}
\usepackage{amsmath,amssymb,amsfonts}
\usepackage{algorithmic}
\usepackage{graphicx}
\usepackage{booktabs} 
\usepackage{textcomp}
\usepackage{xcolor}
\def\BibTeX{{\rm B\kern-.05em{\sc i\kern-.025em b}\kern-.08em
    T\kern-.1667em\lower.7ex\hbox{E}\kern-.125emX}}
\begin{document}
\bibliographystyle{IEEEtran}

\title{Translating Mental Imaginations into Characters with Codebooks and Dynamics-Enhanced Decoding
}

\newcommand{\yansen}[1]{{\bf \color{brown} [[Yansen says ``#1'']]}}
\newcommand{\nielin}[1]{{\bf \color{red} [[nielin says ``#1'']]}}

\author{
\centering
\IEEEauthorblockN{Jingyuan Li\IEEEauthorrefmark{1}\textsuperscript{1},
Yansen Wang\IEEEauthorrefmark{2}\textsuperscript{\textsection},
Nie Lin \IEEEauthorrefmark{3}\textsuperscript{1},
Dongsheng Li \IEEEauthorrefmark{2}}
\IEEEauthorblockA{\IEEEauthorrefmark{1}\textit{University of Washington, Seattle, USA}}
\IEEEauthorblockA{\IEEEauthorrefmark{2}\textit{Microsoft Research Asia, Shanghai, China}}
\IEEEauthorblockA{\IEEEauthorrefmark{3}\textit{The University of Tokyo, Tokyo, Japan}}
}

\maketitle
\begingroup\renewcommand\thefootnote{\textsection}
\footnotetext{Correspondence to Yansen Wang $<$yansenwang@microsoft.com$>$. $^1$The work was done during the internship of Jingyuan Li and Nie Lin at Microsoft Research Asia.}
\endgroup
\begin{abstract}
Advancements in non-invasive electroencephalogram (EEG)-based Brain-Computer Interface (BCI) technology have enabled communication through brain activity, offering significant potential for individuals with motor impairments. Existing methods for decoding characters or words from EEG recordings either rely on continuous external stimulation for high decoding accuracy or depend on direct intention imagination, which suffers from reduced discrimination ability. To overcome these limitations, we introduce a novel EEG paradigm based on mental tasks that achieves high discrimination accuracy without external stimulation. Specifically, we propose a codebook in which each letter or number is associated with a unique code that integrates three mental tasks, interleaved with eye-open and eye-closed states. This approach allows individuals to internally reference characters without external stimuli while maintaining reasonable accuracy. For enhanced decoding performance, we apply a Temporal-Spatial-Latent-Dynamics (TSLD) network to capture latent dynamics of spatiotemporal EEG signals. Experimental results demonstrate the effectiveness of our proposed EEG paradigm which achieves five times higher accuracy over direct imagination. Additionally, the TSLD network improves baseline methods by approximately 8.5\%. Further more, we observe consistent performance improvement throughout the data collection process, suggesting that the proposed paradigm has potential for further optimization with continued use.
\end{abstract}


\section{Introduction}
Brain-Computer Interfaces (BCIs) have revolutionized communication between the brain and external devices, offering transformative solutions for individuals with severe motor impairments \cite{wolpaw2002brain, vidal1973toward}. Among various BCI approaches, non-invasive electroencephalogram (EEG) recording stands out for its safety and practicality, with a long history of use in communication systems \cite{dewan1967occipital, rezeika2018brain}.

\begin{figure}[]
\centerline{\includegraphics[width=0.45\textwidth]{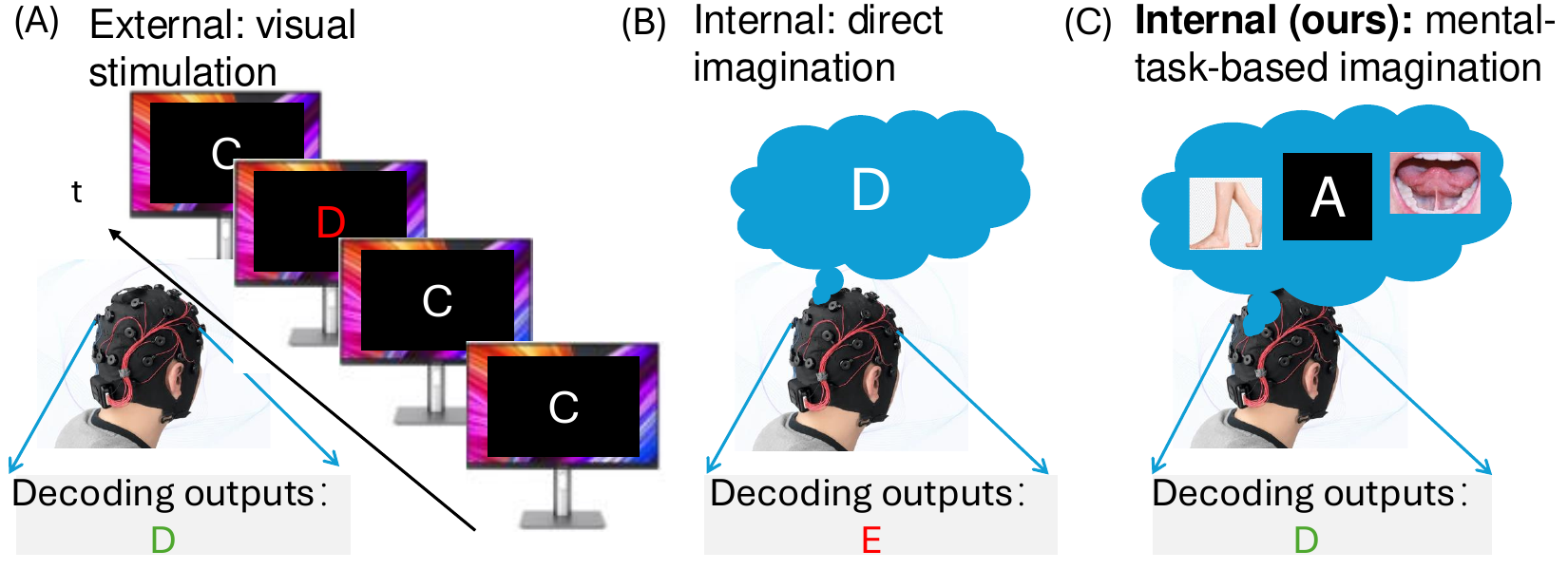}}
\caption{Illustration of EEG signal-to-character mapping paradigms triggered by (A) external stimulation, (B) internal direct imagination, and (C) internal mental-task-based imagination.}
\label{fig-paradigm}
\end{figure}

Recent EEG-based BCI systems for communication  primarily fall into two main categories based on their reliance on external stimuli. Stimulus-dependent systems utilize external cues to evoke Event-Related Potentials (ERPs) like the P300 or Steady-State Visually Evoked Potentials (SSVEP), producing distinct EEG patterns mapped to letters or numbers during decoding \cite{picton1992p300, norcia2015steady}. Combining P300 and SSVEP approaches can enhance decoding performance \cite{nakanishi2014high, chen2014hybrid, nakanishi2017enhancing, bai2023hybrid}. However, these systems require users to focus on external stimuli, often necessitating sustained attention on a screen (see Fig. \ref{fig-paradigm}(A)), which can be unnatural or cumbersome to use as a communication tool.

Alternatively, stimulus-independent systems leverage internal cognitive processes to generate neural activity for word or character selection, offering greater flexibility and user comfort.
Several studies have explored the decoding of vowels, words \cite{coretto2017open, cooney2020evaluation,sarmiento2021recognition}, and even sentences \cite{duan2023dewave}. Nonetheless, these methods are typically limited to small vocabularies, insufficient for daily communication. While decoding full sentences is possible \cite{duan2023dewave}, reliance on techniques like teacher forcing during inference restricts their practical utility.

To address these limitations and enable open-set word generation with high decoding accuracy, we introduce a novel EEG decoding paradigm employing four distinct mental tasks: two types of Motor Imagery (MI), one Visual Imagery (VI) task, and one Arithmetic Computation (AC) task. MI, involving the imagination of body movements and primarily engaging the frontal and parietal lobes \cite{pfurtscheller1997motor}, has been widely used for control signals in cursor control, robotic arm manipulation \cite{forenzo2024continuous, padfield2019eeg}, and BCI spelling \cite{cao2017synchronous, zhang2018converting}. VI tasks activate the parietal and occipital lobes \cite{maghsoudi2021mental, ganis2004brain}, while AC tasks engage the fronto-parietal network \cite{hawes2019neural}. By leveraging mental tasks with distinct neural activation patterns, our paradigm enhances decoding performance. These tasks are mapped to letters and numbers, which can be combined to form words and sentences, enabling open-set word generation.

%

Specifically, we represent each letter or number as a unique sequence of three mental tasks out of four, interleaved with eye-open and eye-closed states to distinguish between tasks. To decode these mental tasks and eye states, we introduce a Temporal-Spatial-Latent-Dynamics (TSLD) network.  Although mental-task-based imagination is less intuitive than direct imagination, it achieves approximately five times higher accuracy using a comparable amount of data and similar model architectures. Besides, the proposed TSLD network further enhances decoding accuracy by around 8\% compared to baseline methods. We additionally observe that decoding performance improves progressively throughout the data collection process. These results highlight the potential of combining mental tasks for EEG-based open-set communication.

\section{Methods}
\begin{figure}[]
\centerline{\includegraphics[width=0.48\textwidth]{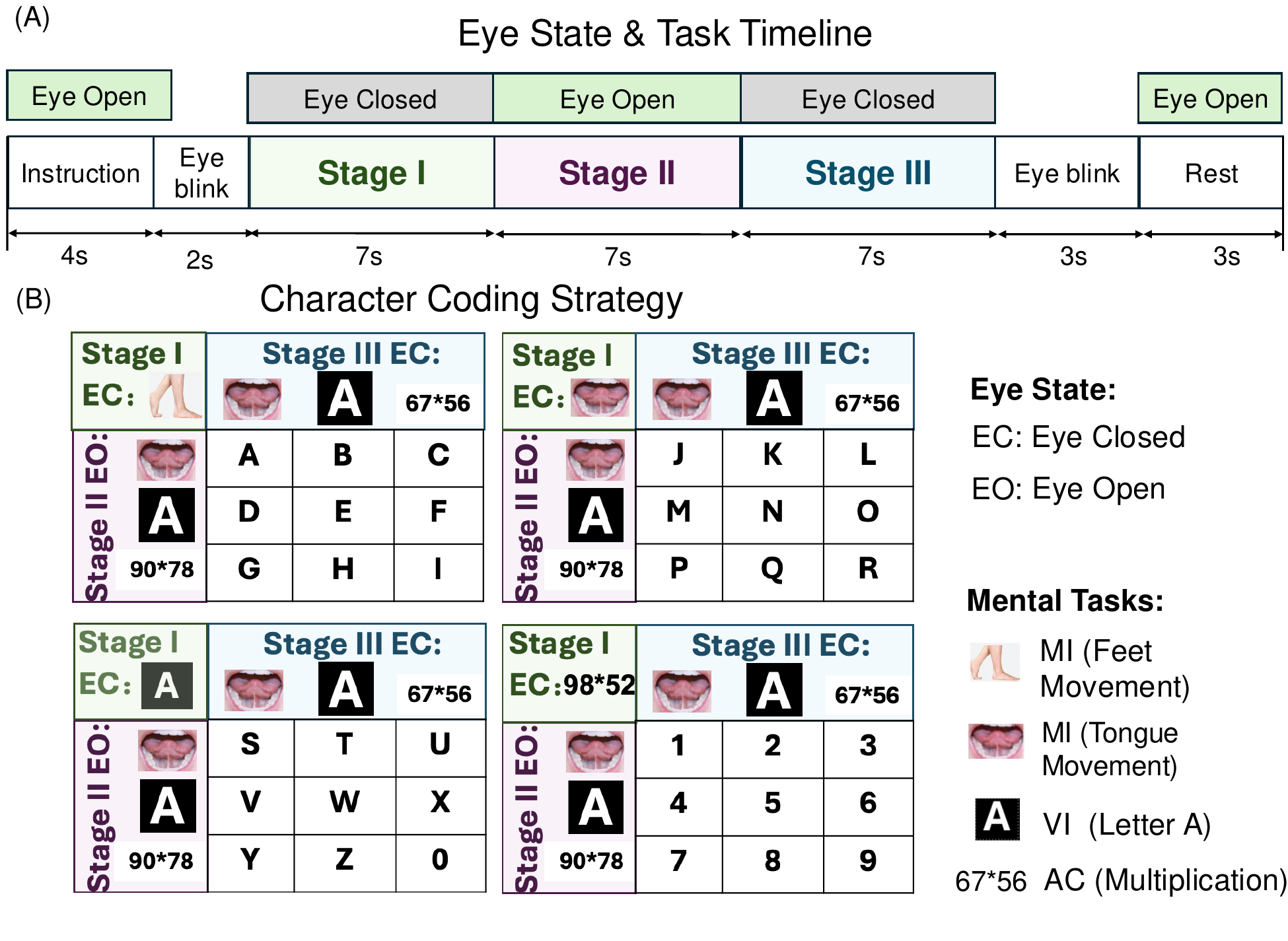}}\caption{(A)Data collection timeline illustrating a sequence of three mental task stages interleaved with eye-open and eye-closed states. (B)  Brain-to-Character codebook mapping characters to specific combinations of mental tasks.}
\label{fig-coding}
\end{figure}
\subsection{Coding paradigm}
Several methods exist for mapping EEG recordings to the characters an individual intends to communicate. High-accuracy decoding systems typically rely on external stimuli to evoke specific brain responses, such as the P300 signal \cite{fazel2012p300}, as shown in Fig. \ref{fig-paradigm}(A). A more natural approach, which does not require external stimuli, involves direct imagination (Fig. \ref{fig-paradigm}(B)), where individuals imagine characters, and the system decodes these from the EEG signals. While intuitive, this approach often suffers from performance limitations.

To allow individuals to select letters freely without reliance on external stimuli, we propose a novel coding paradigm where each character is represented by a unique sequence of three consecutive mental tasks. These tasks are chosen from Motor Imagery (MI) of foot or tongue movement, Visual Imagery (VI) of the letter 'A' on a white background, and Arithmetic Computation (AC) involving the multiplication of two-digit numbers. The selected tasks are designed to activate distinct brain regions, thereby optimizing discrimination capability \cite{pfurtscheller1997motor, ganis2004brain, hawes2019neural}. Individuals perform these specific mental tasks, and the corresponding EEG signals are decoded to identify the intended character (Fig. \ref{fig-paradigm}(C)). 

We illustrate the unique mappings from mental tasks to characters in Fig.\ref{fig-coding}(B), with each character associated with three stages of mental tasks alternating between eye-open and eye-closed condition. For instance, in reference to letter 'D', the subject first imagines foot movement with eyes closed, then imagines tongue movement with eyes open, followed by imagining the letter 'A' with eyes closed. Eye blinks mark the beginning and end of each imagination phase. This system can be seamlessly extended to multiple characters by adding additional sequences of three-stage imaginations.

\begin{figure*}[]
\centerline{\includegraphics[width=0.9\textwidth]{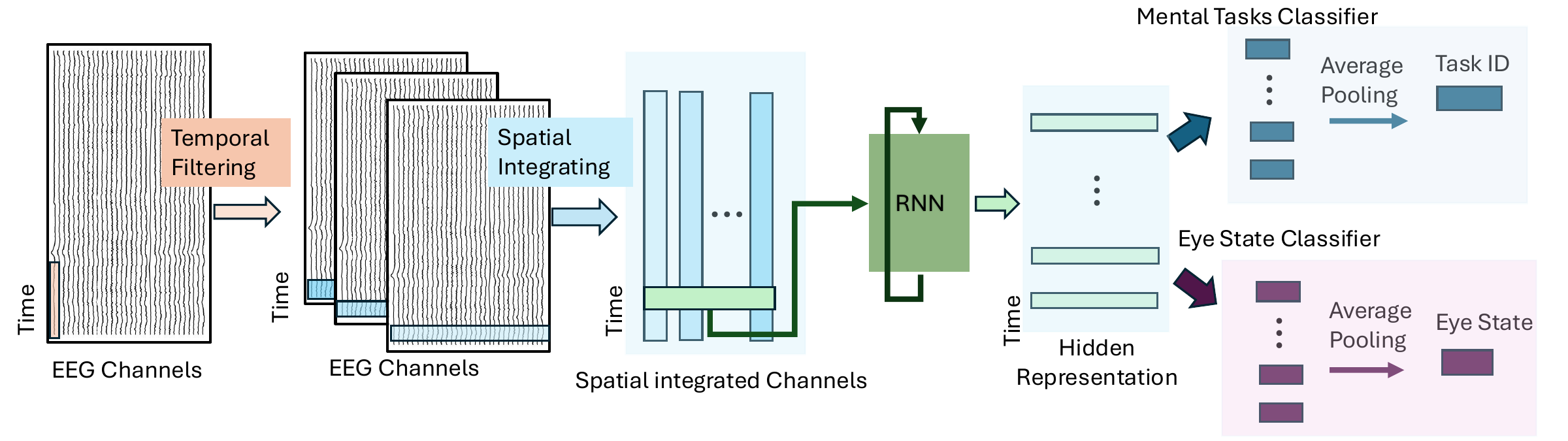}}
\caption{Overview of Temporal-Spatial-Latent-Dynamics (TSLD) architecture.}
\label{fig-architecture}
\end{figure*}
\subsection{Data Preprocessing}
EEG recordings capture both neural activity related to mental tasks and unwanted artifacts, such as muscle movements, eye movements, and slow drifts \cite{jiang2019removal, de2018robust}. These artifacts can contaminate the signal of interest and interfere with subsequent analysis. To reduce these effects, we implement several preprocessing steps.
First, we detrend the EEG signals to remove slow drifts \cite{de2018robust}. Next, we apply a band-pass filter between 4 Hz and 38 Hz, removing low-frequency signals (below 4 Hz) associated with eye movements, as well as high-frequency noise above 38 Hz \cite{schirrmeister2017deep}. The signals are then re-referenced using common average referencing resulting the spatotemporal EEG matrix, 
$\hat{X} \in \mathcal{R^{T \times C}}$
 , where  $T$
denotes the timesteps, and 
$C$ represents the number of EEG channels. To ensure model training stability, we perform exponential moving standardization on  
$\hat{X}$ \cite{10.3389/fnins.2013.00267, HBM:HBM23730}.
Specifically, we compute moving average at time t, $m_{[t,i]}= \frac{ \sum_{j\in [0,t]}(1-\alpha)^j\hat{X}_{[t-j, i]} }{(1-\alpha)^j}$, and corresponding standard deviation, $ v_{[t,i]}= \frac{\sum_{j\in [0,t]}(1-\alpha)^jm_{[t, i]}^2 }{(1-\alpha)^j}$.
The the normalized EEG signal is computed as $ X_{[t,i]} = \frac{(X_{[t,i]} - m_{[t,i]})}{max(\sqrt{v_{[t,i]}}, \epsilon)}$ and is further taken as inputs to TSLD to predict mental task categories and eye states.


\subsection{Model Architecture}
Mental tasks such as MI, VI, and AC activate distinct brain subnetworks, each operating across different frequency bands \cite{maghsoudi2021mental,ganis2004brain, hawes2019neural,pfurtscheller1997motor}.
 However, EEG data also contains physiological artifacts like cardiac and muscular activity, as well as external noise \cite{biasiucci2019electroencephalography}, complicating the isolation of task-related signals. 
To address this, we propose the Temporal-Spatial-Latent-Dynamics (TSLD) network, which isolates task-relevant signals through three key steps: (i) temporal filtering for signal decomposition, (ii) spatial integration to extract features from relevant brain subnetworks, and (iii) dynamic signature discovery using a Recurrent Neural Network (RNN) to capture temporal dependencies.

In the signal decomposition step, we apply temporal convolution independently for each channel, which filters EEG signal into different frequency components. These frequency components are learnable through temporal convolution kernel weights and have been used in previous studies \cite{HBM:HBM23730, lawhern2018eegnet},  as opposed to predefined bandpass filters for decomposition \cite{ang2008filter}.  
Specifically, the original signal $X \in \mathcal{R}^{C \times T}$ is transformed into $X' \in \mathcal{R}^{C \times F \times T'}$, where $F$ is the number of output channels in temporal convolution layer deciding the number of learned frequency components.
Then the decomposed signals from $C$ EEG channels and $F$ frequency bands are integrated to obtain subnetwork representation independently for each timesteps. Here, we use spatial convolution for integration \cite{HBM:HBM23730}, resulting in subnetwork represntation of dimension $F' \times T'$, where $F'$ is the number of output channels reflecting the number of subnetworks.
In the following step, a Recurrent Neural Network (RNN), specifically Gated Recurrent Units (GRU) \cite{chung2014empirical}, process extracted subnetwork representations at each time step.
The GRU captures temporal dependencies and generates latent outputs that encapsulate dynamic information for each time step. Then the mental task classifier and the eye state classifier independently categorize the latent dynamics into mental task and eye state. The final class label is determined by average pooling over the entire temporal sequence.
The TSLD network are optimized by minimizing both the eye state loss ($\mathcal{L}_{es}$) and the mental task loss ($\mathcal{L}_{mt}$) simultaneously. $\mathcal{L}_{es}$ and $\mathcal{L}_{mt}$ are standard cross entropy loss with 2 categories for eye state classification and 4 categories for mental tasks. 

\subsection{Model Training and Inference}
During model training, the TSLD model randomly selects a 1000-timestep EEG segment as input, from which it estimates both the mental task category and the eye state. The model is optimized by minimizing $\mathcal{L}_{es}$ and $\mathcal{L}_{mi}$ for every possible 1000-step window. 
During inference, the TSLD model processes full segments of 1792 timesteps all at once. It generates predictions for the eye state and the mental task category within each 1000-timestep window, shifting the window by 100 timesteps at a time. These overlapping predictions across the full segment are then aggregated to compute adjusted class probabilities and make the final prediction over the entire segment. Multiple methods can be used to compute the adjusted probabilities and, in practice, majority voting yields the best performance. In particular, the probability for each class is determined by the proportion of time windows assigned to that class in the full sequence. A target character is considered correctly predicted only if both the mental task categories and the eye states are accurately recognized across all three stages for a particular character.

\section{Experiments}
\begin{figure}[]
\centerline{\includegraphics[width=0.38\textwidth]{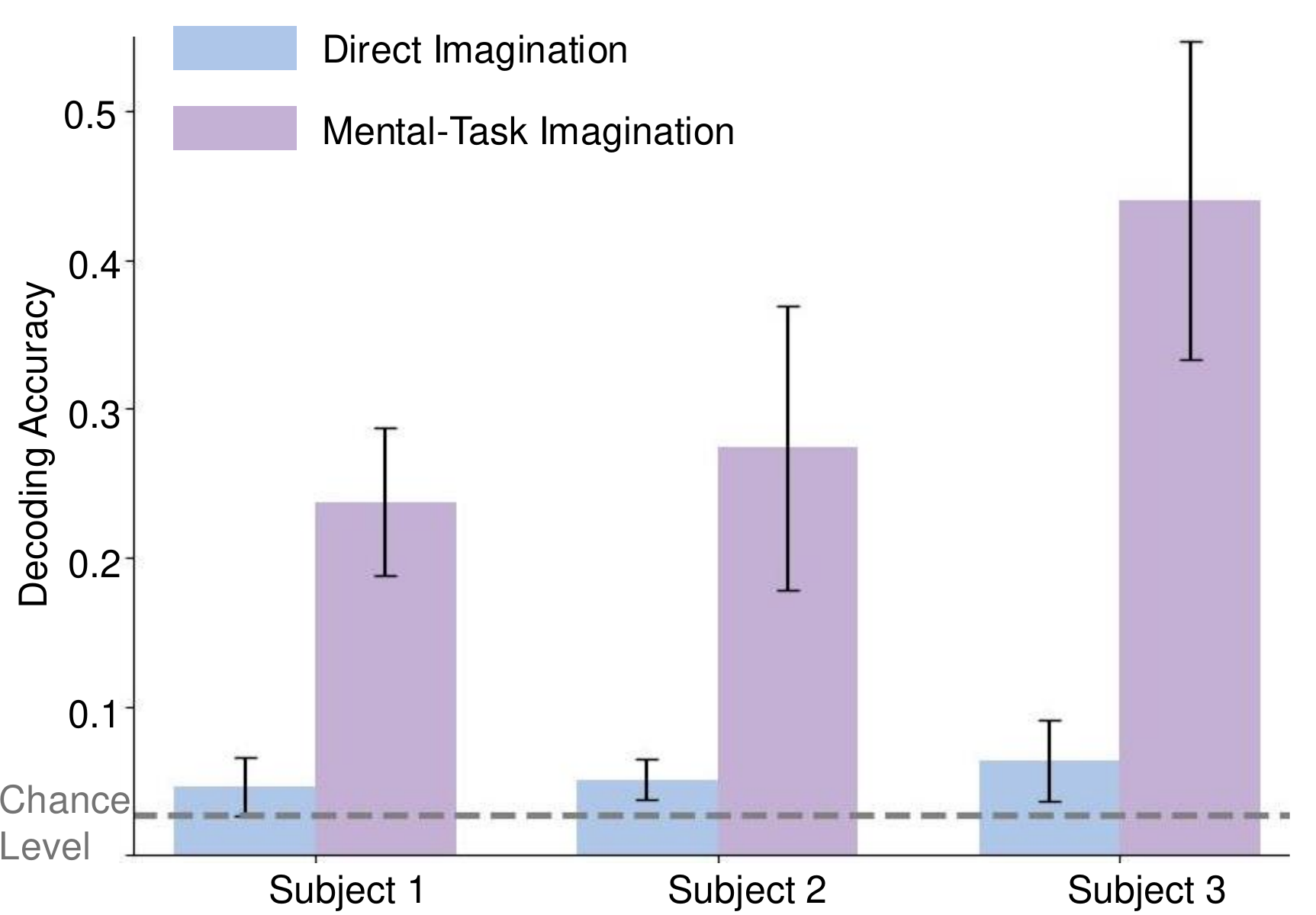}}
\caption{Comparison of performance between direct imagination and mental-task-based imagination, with mental-task-based imagination being around five times more accurate than direct imagination.}
\label{fig-direct-related-comparison}
\end{figure}

\begin{figure}
    \centering
    \includegraphics[width=0.48\textwidth]{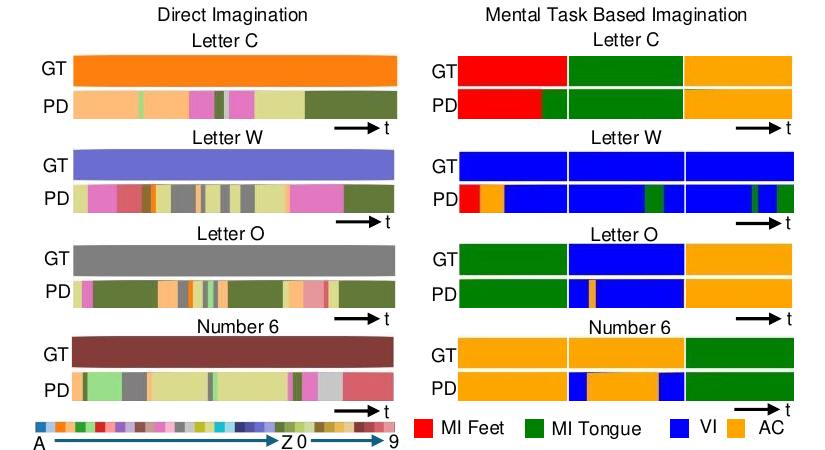}
    \caption{Example characters of ground truth (GT) and model prediction (PD) in full imagination window using the direct imagination paradigm (left) and the mental-task-based imagination paradigm (right). PD with mental-task-based imagination is more coherent and accurate than PD with the direct imagination paradigm.}
    \label{fig:decoding-examples}
\end{figure}

\subsection{Data Collection}
To evaluate the effectiveness of the proposed coding paradigm, we collecte EEG recordings using the Emotive FLEX 2 Gel 
32-Channel Wireless EEG headcap system at a sampling rate of 256 Hz. Participants are instructed to perform mental tasks corresponding to each character as defined by the proposed codebook (Fig.\ref{fig-coding}(B)). The timeline for collecting EEG signals for a single character is detailed in Fig.\ref{fig-coding}(A). At the beginning of a character associated imagination, an audio cue indicates the target character, followed by a 4-second display of the codebook to help participants identify the corresponding mental tasks. Eye blinks mark the start and end of each imagination process. Between these blinks, participants execute a specific sequence of three mental tasks to reference the target character.

Three subjects participated in the data collection. Each subject completed six experimental sessions involving mental-task-based imagination for all 36 characters, presented in random order. EEG recordings from the three-stage imagination time window are used for TSLD training.
In a separate set of experiments, participants were asked to directly imagine the 36 characters. For consistency, each participant performed six additional experimental sessions, imagining all 36 characters in random order for 21 seconds each, matching the duration of the three-stage mental-task-based imagination.

For training and validation, we employ a six-fold cross-validation strategy, with five out of the six sessions for training and the remaining one for validation.


\subsection{Performance Comparison on Imagination Paradigms}
In this section, we quantitatively compared character decoding accuracy between two paradigms: direct imagination and the proposed mental-task-based imagination. For both paradigms, the TSLD model is trained to make predictions over a 1000-timestep window. Notably, for direct imagination, the eye state classification branch of TSLD is removed, and the model is trained as a 36-way classifier.
During inference, we apply a sliding window approach, making predictions on the full sequence for every 1000 timesteps with a 100-step window shift. Adjusted probabilities are then computed to generate the final predicted classes for both direct and mental-task-based imagination. The decoding performance across the three subjects is shown in Fig.\ref{fig-direct-related-comparison}, where purple represents the decoding accuracy for mental-task-based imagination, and blue represents direct imagination. The error bars reflect results from six-fold cross-validation with three random seeds.
The results clearly demonstrate the superiority of the mental-task-based imagination paradigm, with accuracy nearly five times higher than that of direct imagination. This highlights the significant advantage of using the proposed mental-task-based character codebook for BCI systems.

We also visualize the predicted (PD) mental task categories and the predicted direct imagination categories throughout the full imagination duration, alongside the ground truth (GT), in Fig.\ref{fig:decoding-examples}. While there may be occasional mistakes in the decoding of mental tasks across whole imagination stages, the overall accuracy is maintained. In contrast, direct imagination generally produces more erratic decoding outputs. 



\subsection{Performance Comparison across Model Architecture}

\begin{table}[]
\caption{Decoding Accuracy on Different Model Architecture}
\begin{center}
\begin{tabular}{cccc}
\toprule
 & \textbf{Top1 ACC}& \textbf{Top3 ACC}& \textbf{Top5 ACC} \\
\midrule
EEGV4 \cite{lawhern2018eegnet}                         & 0.2758                        & 0.4745 & 0.5815 \\
ShallowConvNet \cite{HBM:HBM23730}                       & 0.2744                         & 0.4826 & 0.5771 \\
  ATCNET \cite{altaheri2022physics}                     & 0.2955                      & 0.5034 & 0.5898 \\
EEGConformer \cite{song2022eeg}                 & 0.2961                         & 0.4878 & 0.5727 \\
\midrule
\textbf{TSLD (ours)}                         & \textbf{0.3204}                         & \textbf{0.5312} & \textbf{0.6128} \\
\bottomrule
\end{tabular}
\label{tab2: model architecutre}
\end{center}
\end{table} 
We further evaluated several well-established models on the mental-task-based imagination task. Specifically, we compared the following models: ShallowConvNet \cite{schirrmeister2017deep}, a one-layer spatial-temporal graph convolutional network; EEGNetV4 \cite{lawhern2018eegnet}, a compact, multi-layer convolutional neural network; ATCNET \cite{altaheri2022physics}, which incorporates an attention block with sliding window concatenation and temporal convolution layers for EEG signal representation; and EEGConformer \cite{song2022eeg}, which appends a Transformer after convolutional patching of the input EEG signals.
For each method, we modified the original single-classifier architecture to a two-classifier setup, predicting both eye states and mental task categories. The decoding performance of these models, averaged across three subjects, is presented in Table \ref{tab2: model architecutre}, including Top 1, Top 3, and Top 5 accuracy. The results show that
our proposed method consistently outperforms the baseline models, emphasizing the importance of capturing dynamic EEG signal patterns across different subnetworks to accurately identify the mental tasks being executed. These results further validate the effectiveness of the TSLD architecture in improving decoding accuracy for mental-task-based imagination tasks.

\subsection{Decoding Performance Improves Through Learning}
\begin{figure}[]
    \centering
    \includegraphics[width=0.47\textwidth]{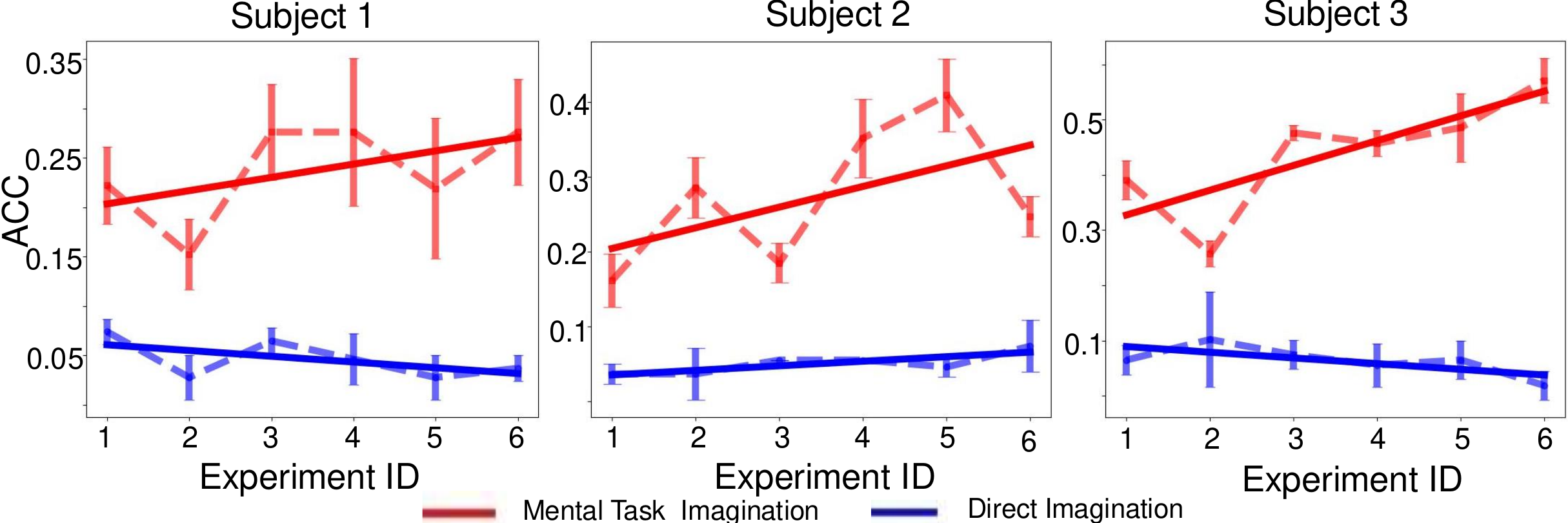}
    \caption{Decoding accuracy across six experiments.}
    \label{fig:decoding-acc-6sessions}
\end{figure}

To be noticed, the decoding accuracy of mental-task-based can be improved with continued data collection procedure. In particularly, we find for all three subjects, the decoding accuracy improves at the later stage of data collection. As shown in Fig \ref{fig:decoding-acc-6sessions}, 
there is a consistent trend of improving the decoding performance along the data collection procedure for all three subjects during mental-task based imagination (red curve). However, no significant decoding accuracy changes observed during the direct imagination (blue curve). 


\section{Conclusion}
In this work, we present a novel mental-task-based coding paradigm and a temporal-spatial-latent-dynamics (TSLD) network that accurately maps EEG recordings to intended characters without relying on external stimuli. The observed gradual improvement in decoding performance throughout the data collection procedure highlights the potential for further enhancing accuracy with the mental-task-based imagination paradigm in future BCI applications.

\bibliography{ms}

\end{document}